# Numerical investigation of the flow induced by a transcatheter intra-aortic entrainment pump

Yeojin Park[a], Osman Aycan[a,b], Lyes Kadem[a]

[a]Laboratory of Cardiovascular Fluid Dynamics, Department of Mechanical Industrial and Aerospace Engineering, Concordia University, Montreal, QC, Canada; [b]Department of Mechanical Engineering, Faculty of Engineering, Zonguldak Bulent Ecevit University, Zonguldak, Türkiye

**Correspondence:** Osman Aycan, Email: osman.aycan@beun.edu.tr; lcfd@encs.concordia.ca.

## ABSTRACT

This study evaluates the fluid dynamics inside and outside transcatheter blood pump positioned in the aorta. We focus on the pump's impact on blood component damage and arterial wall stress. CFD simulations were performed for rotational speeds ranging from 6000 to 15000 rpm, with a blood flow rate of 1.6 L/min. Results show that significant blood damage may occur at speeds as low as 12000 rpm, and the pump's outflow jet induces elevated wall shear stress, potentially leading to arterial aneurysms. These findings suggest the need for further design improvements to reduce risks when used in prolonged or transplant-related applications.

## 1. Introduction

Heart failure (HF) is a pathological condition where the heart is unable to supply enough blood to the rest of the body. Patients with end-stage heart failure that are refractory to medical therapy have an extremely high mortality rate that can reach up to 80% in 1 year. The most viable option in such patients is heart transplantation. However, the shortage of heart donors leads to an ever-growing patient waiting list. Under such conditions, left ventricular assist devices (LVADs) represent the only alternative for patients with heart failure. The main role of LVADs is to help manage heart failure by assisting blood circulation from the ventricle to organs and tissues (Lima et al. 2015). However, several patients may not necessarily be eligible for long-term LVAD support, as this might significantly impact their quality of life (Meyns et al. 2011; Lu et al. 2022). In such patients, a more optimal approach might consist in a short-term use of continuous flow percutaneous mechanical circulatory support that can promote cardiac rest and support heart remodeling, contributing to a reduction in the risk of morbidity and mortality and serving as a good bridge to recovery.

The current devices are mostly positioned at the level of the ascending aorta across the aortic valve to directly support the failing heart. However, such positioning can lead to elevated risks of stroke (Tang et al. 2023). This led to the development of alternative designs and approaches including positioning the percutaneous mechanical circulatory support in the descending aorta. Recent studies have shown that optimizing flow conditions in axial blood pumps can significantly mitigate complications like hemolysis and thrombosis, which are critical in high-risk patients (Zhu et al. 2024). The reported anticipated benefits of such positioning consist of avoiding the interaction with the native aortic valve and function and minimalizing the risk of thromboembolic stroke (Georges et al. 2022). Another reported anticipated benefit is an increase in the renal artery blood flow (Grafton et al. 2023; Nathan and Basir 2023).

Such intra-aortic entrainment pumps are currently at different stages in their design and validation, with some in the early phases and others successfully implanted in humans with ongoing trials, requiring therefore more in-depth investigations. To the best of our knowledge, Li et al. (2022) conducted the first study investigating the impact of intra-aortic entrainment pumps on flow dynamics in the descending aorta. Using computational fluid dynamics simulations, they demonstrated that positioning the pump near the suprarenal abdominal aorta optimizes the improvement of flow dynamics in the aorta. Additionally, they showed that increasing the number of pumps can lead to a satisfactory flow rate at a lower rotational pump speed, thereby reducing energy loss and shear stress on blood particles. Despite the significance of these findings, it is important to note that the authors did not consider a realistic scenario in which a physiological pulsatile flow from the heart interacts with the continuous flow generated by the intra-aortic pump. In this context, it is interesting to note that intra-aortic entrainment pumps pose a new interesting and mostly unexplored challenge since the pump significantly interacts with patient's anatomy and residual pulsatile hemodynamic conditions emanating from the failing heart. Under such conditions, investigating the flow dynamics inside and outside the pump and the impact of the pump outflow on the aortic wall becomes crucial since simply simulating an isolated pump and focusing solely on the internal flow, as commonly performed in the literature, does not provide enough information on the pump's behavior. Furthermore, it has been shown that variations in impeller axial position in centrifugal blood pumps significantly impact hemodynamics and hemolysis outcomes, indicating that design optimization is essential for reducing blood damage (Lv et al. 2024).

In this study, we use computational fluid dynamics (CFD) to investigate, the flow dynamics, risks of hemolysis and changes in arterial wall shear stress when a model of an intra-aortic entrainment pump is positioned in a realistic anatomy of a descending aorta and ran at different rotational speeds.

## 2. Methods

### *2.1. Device and aorta geometry*

In this study, the pump geometry was obtained by a reverse engineering version of the HeartMate 2 geometry with reference to Blum et al. (2022) and Romanova et al. (2022). The initial geometry was modified by considering axial flow mechanical circulatory support devices such as Aortix$^{TM}$ (Procyrion, Houstan, TX, US) and ModulHeart$^{TM}$ (Puzzle Medical Devices, Montreal, QC, Canada). The modification of the initial geometry was carried out according to the pre-simulations, and the final design parameters were obtained as shown in **Figure 1**. However, it provides a great potential for future studies in terms of improving and optimising the design used in our study as a result of more detailed numerical analysis results to be obtained. . Here it is important to note that the objective of this study is not to propose a new design for an intra-aortic entrainment pump, but rather to simulate the impact of such devices on flow dynamics and the aortic wall. The design process was performed using CAD software, SolidWorks (Dassault Systèmes, France). The axial blood pump is suitable for intravascular usage, and it consists of a flow straightener, an impeller, and a diffuser. The detailed design features include a body diameter of 7.6 mm for the impeller part and an impeller diameter with blades of 9 mm. The clearance between the diffuser blades and pump cover body is 0.2 mm. The pump cover is a protective cage for the device in the artery. The diffuser blade angle is 10.3°, and the impeller blade angle is 43.5°. The intra-aortic entrainment pump has a body with a 10 mm diameter over the pump blades and a motor part with the same dimensions towards the outflow. More detailed dimensions (in mm) and blood pump geometry are shown in **Figure 1c**.

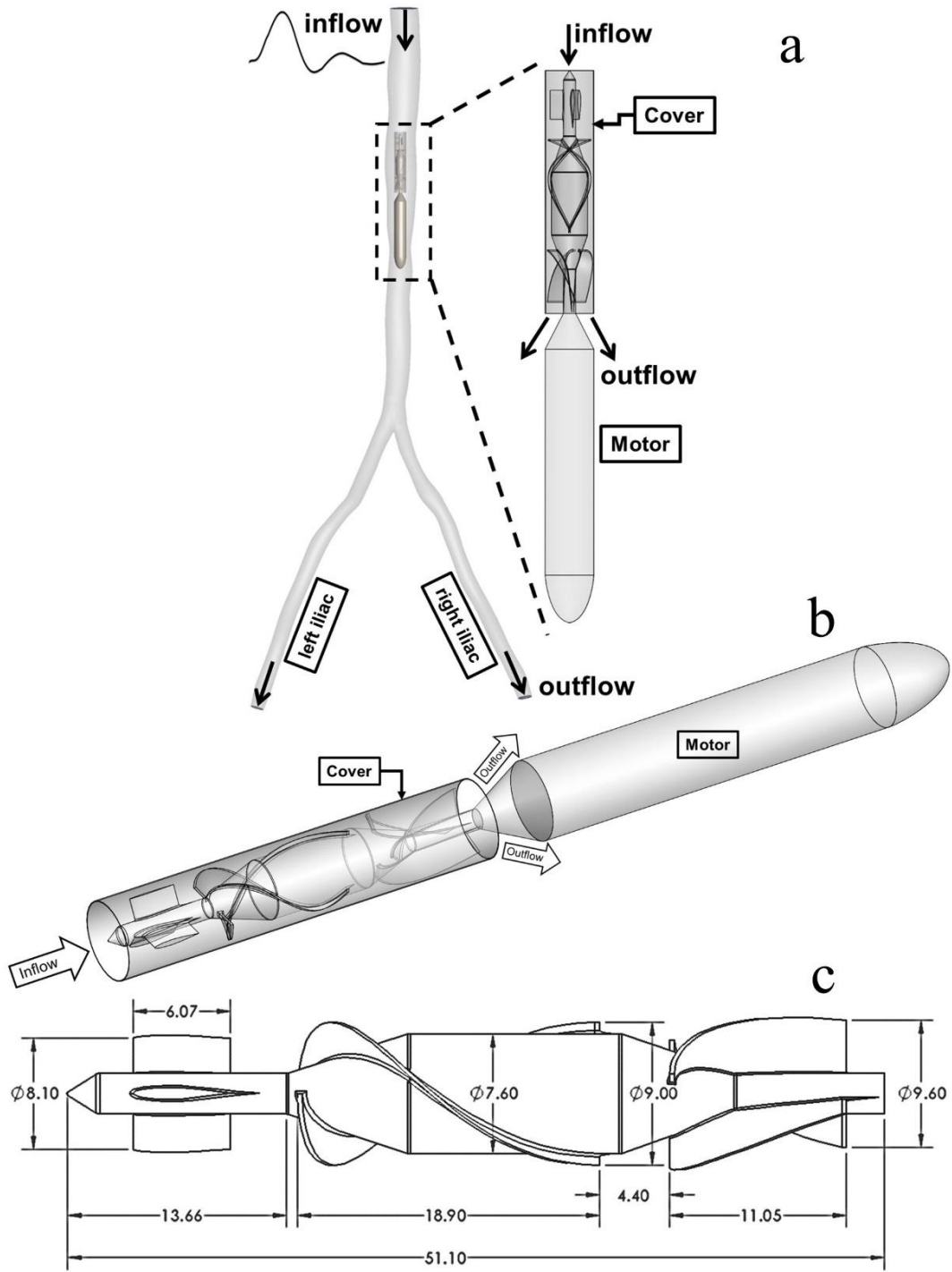

**Figure 1.** (a) Design features of transcatheter intra-aortic blood pump and its positioning inside the patient-specific abdominal aorta; (b) Isometric view of the transcatheter intra-aortic blood pump; (c) The dimensions of the pump body and blades.

In this study, computational fluid dynamic simulations have been conducted by implementing the intra-aortic entrainment pump inside a patient-specific abdominal aorta geometry as seen in **Figure 1a**. The patient-specific aorta's CAD model was generated from DICOM data sourced from the open-source database (Lan et al. 2018). The diameter of the aorta at the entrance is 25.07 mm, and the left and right iliac diameters at the outlet are 11.17 mm and 12.06 mm, respectively.

*2.2. Domain meshing*

The pump and aorta models were discretized using polyhedral elements and prism layers to ensure a y+ value of less than 1. The advantage of using polyhedral elements has been highlighted in a previous study (Aycan et al. 2023). This discretization was performed using Fluent Meshing v22 (ANSYS Inc., Canonsburg, PA). The computational domains contain volumetric elements ranging in size from 0.14 mm to 2.00 mm 0.41 mm to 2.00 mm inside aorta and 0.14 mm to 0.65 mm inside the pump cover while the size of the surface elements on the pump geometry ranges from 0.041 mm to 0.2 mm. For boundary layer resolution, 5 layers were employed with a high near-wall resolution. This approach enabled the use of a low Reynolds formulation of the turbulence model, specifically the *k-ω* shear stress transport model (Schüle et al. 2016). Various grid sizes for polyhedral elements were decreased with a refinement factor of 1.3 to increase the mesh density (Celik and Zhang 1995), leading to several elements ranging from 0.67 to 10.4 million across the entire domain. **Table 1** provides information about the cell sizes and the number of elements, and a detailed visualization of the computational mesh for the pump is shown in **Figure 2**. All mesh cases in this study have no negative volumes. The maximum aspect ratio across all cases was 14.5, and the minimum orthogonal quality was 0.22.

**Table 1.** Overview of the generated mesh for the entire geometry.

|  | Cell Sizes | | | Number of Layers | | | Cell Numbers |
|---|---|---|---|---|---|---|---|
|  | *Surface (Pump Surface) [mm]* | *Volume (Aorta) [mm]* | *Volume (Pump Region) [mm]* | *Near-Wall (Aorta)* | *Near-Wall (Inner Cover)* | *Near-Wall (Outer Cover and Motor)* | *Polyhedral Elements* |
| Case 1 | 0.200 | 2.00 | 0.65 | 5 | 4 | 5 | 673 382 |
| Case 2 | 0.154 | 1.54 | 0.50 | 5 | 4 | 4 | 924 036 |
| Case 3 | 0.118 | 1.18 | 0.38 | 5 | 4 | 4 | 1 122 124 |
| Case 4 | 0.091 | 0.91 | 0.30 | 5 | 4 | 4 | 1 699 706 |
| Case 5 | 0.070 | 0.70 | 0.23 | 5 | 4 | 4 | 3 035 089 |
| Case 6 | 0.054 | 0.54 | 0.18 | 5 | 4 | 4 | 5 450 633 |
| Case 7 | 0.041 | 0.41 | 0.14 | 5 | 4 | 4 | 10 448 443 |

The simulation was performed over three cardiac cycles, disregarding initial effects as described in Section 2.3. The shear stress tensors ($\tau_{ij}$) were collected from the last cardiac cycle, and time-averaged values were considered to be able to calculate the scalar shear stress by using Bludszuweit's stress formula for mesh independence study as explained in Section 2.5. One of the most important parameters is the hemolysis index (HI) to evaluate the usefulness of the blood pumps. The scalar shear stress plays an important role in the calculation of the HI. In this study, the main focus is to evaluate the effect of the rotational speeds on blood damage (hemolysis index) as well as flow effect at the pump outflow over the aorta wall. Therefore, the shear stress was chosen as the only parameter for the study of mesh independence.

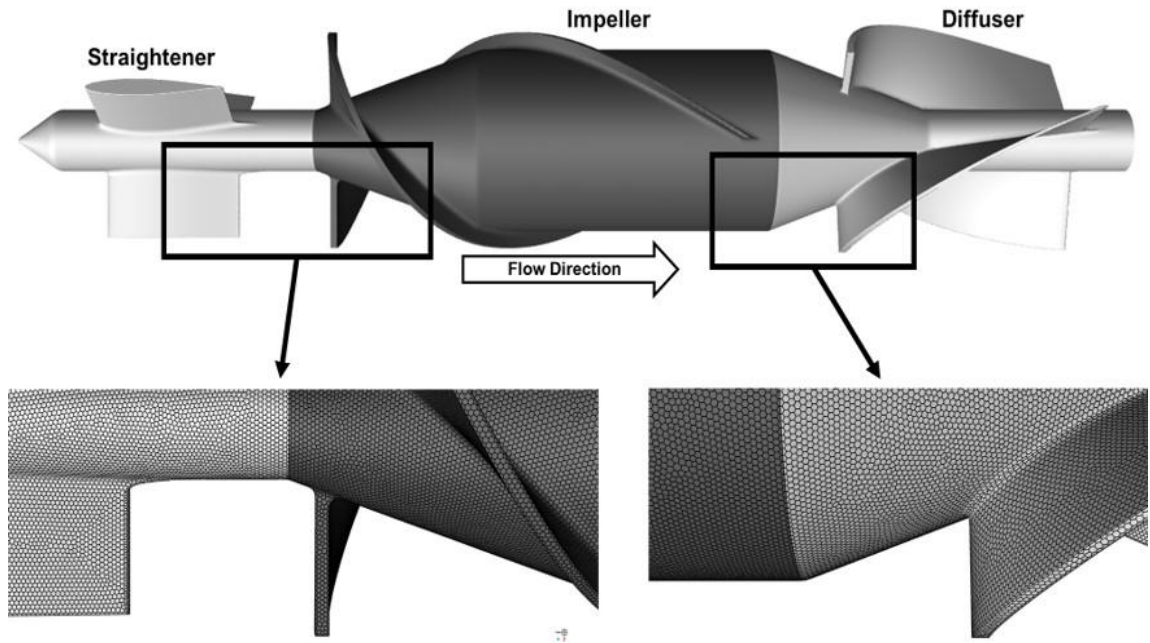

**Figure 2.** Meshing details of intravascular blood pump for RANS simulations.

Mesh independence analyses were conducted by considering shear stress values for the fluid domain inside the pump cover with seven distinct cases, refining the mesh cell sizes to determine an appropriate mesh size that balances accuracy and computational cost. Due to the pump rotation, the most severe deformation rates occur in the fluid domain inside the pump cover in comparison with the other regions. For the sake of robustness, we chose to focus the mesh independence study and GCI analysis on what we consider the worst-case scenario. The difference shows that the difference for each case is below 5%. However, further analysis is still required to determine the most appropriate grid size. The Grid Convergence Index (GCI) Method was employed to evaluate numerical uncertainties resulting from spatial discretization and iterative convergence errors by using the finest five cases (i.e., cases from 3 to 7). All cases have reached the asymptotic range of convergence. The actual fractional error has been calculated depending on the reference shear stress value (9.66 Pa) by using Richardson extrapolation $f_{h=0}$ (Richardson 1911), and the error band is below 10% for case 6 and case 7. Consequently, case 6 with 5.45 million elements has been selected and used for the further simulations.

*2.3. Numerical scheme and boundary conditions*

The simulations were conducted using ANSYS Fluent v22.1 to numerically solve the Reynolds-Averaged Navier-Stokes (RANS) equations. Unsteady simulations were performed utilizing the standard *k-ω* turbulence model. The suggested standard parameters have been kept for the selected turbulence model. The pressure-velocity coupling was achieved using the PISO algorithm, which was applied to solve the momentum and pressure-based continuity equations. To discretize the momentum equations, a second-order upwind scheme was utilized. The convergence criteria for both continuity and velocity were met when the residuals fell below $10^{-4}$ residual.

Again, a unique aspect of this study is its consideration of the pump under realistic conditions, including its location within the abdominal aorta and its interaction with the residual aortic flow waveform. A generic physiological flow waveform (average flow rate of 1.6 L/min; heart rate: 60 bpm) was applied as an inlet boundary condition to the aorta model. To maintain consistency with flow rate conditions that may occur in heart failure patients requiring the use of VADs, the inlet boundary condition corresponds to 45% of the normal resting flow rate of a healthy individual (Pepine et al. 1978). The simulation was conducted over three cardiac cycles, disregarding initial effects. An interface condition was assigned to the inlet and outlet of the blood pump to enable blood flow through it. In **Figure 2**, the pump's straightener and diffuser (light grays) were designated as non-rotating-wall boundary conditions, while the impeller (dark gray) was defined as a rotating-wall boundary condition. The pump's rotational speed varied between 6000 to 15000 rpm. Additionally, the time steps were chosen approximately as 0.140, 0.090, 0.070, and 0.055 ms to result in a 5° pump rotation (Blum et al. 2022) when operating at 6000, 9000, 12000, and 15000 rpm, respectively.

*2.4. Blood rheology*

Blood was used as the working fluid. The rheological behavior of blood is influenced by various factors, including shear rate, domain diameter, temperature, hematocrit, and more. Some research studies suggested that at shear rates exceeding 100 s$^{-1}$, blood exhibits Newtonian fluid characteristics (Hegde et al. 2021).

Nevertheless, relying solely on the assumption of constant blood viscosity may prove inadequate in certain situations, as highlighted by Doost et al. (2016). Therefore, blood rheology is considered in this study by using the Carreau model (Giersiepen et al. 1990). The governing equation for viscosity based on the Carreau model has been expressed as follows:

$$\mu = \mu_\infty + (\mu_0 - \mu_\infty)[1 + (\lambda\dot{\gamma})^2]^{\frac{n-1}{2}} \quad (1)$$

The parameters for this model are derived from the study conducted by Cho and Kensey (1991), where they fitted **Equation 1** with their experimental data. The high-shear viscosity ($\mu_\infty$) and low-shear viscosity ($\mu_0$) values are taken as 0.00345 Pa·s and 0.056 Pa·s, respectively. For the simulations, the time constant ($\lambda$) is set to 3.313 s, and the power-law index ($n$) is chosen as 0.3568. A pressure-outlet condition was applied for the outlet boundary condition, with atmospheric pressure and operating pressure of 100 mmHg. For simplicity and considering that patients with heart failure have a significantly impaired arterial compliance (Desai et al. 2009), the walls of the entire geometry were considered rigid and subject to a no-slip boundary condition.

*2.5. Computational estimation of blood damage*

Hemolysis, a critical stage characterized by the rupture of red blood cells (RBCs), is a significant concern when dealing with intravascular blood pumps. The primary cause of hemolysis is the excessive shear stress exerted on the RBCs, leading to their damage. But, to assess the risk of blood damage, two crucial parameters should be considered: the level of shear stress (both viscous and turbulent) and the exposure time to these shear stresses. A specific approach was employed to calculate the viscous and turbulent shear stresses. This method determines scalar stress, representing the overall stress experienced by the blood, and encompasses the six components of the stress tensor (Cheng et al. 2021). The scalar shear stress for a three-dimensional fluid domain is calculated using Bludszuweit's stress formula (Bludszuweit 1995):

$$\tau_{scalar} = \left[\frac{1}{6}\sum_{i \neq j}(\tau_{ii} - \tau_{jj})^2 + \sum_{i \neq j}\tau_{ij}^2\right]^{\frac{1}{2}} \quad (2)$$

The fluid residence or exposure time is determined by assessing the blood streamlines (Bludszuweit 1995; Kafagy et al. 2015). In this study, 500 particle streamlines are performed to estimate the exposure time and the individual scalar shear stress they experience is computed using **Equation 2**.

The hemolysis model employed in this study was proposed by Giersiepen et al. (1990), commonly known as the power-law relation. This model describes the extent of hemolysis using a power-law function, which considers the magnitude of shear stress ($\tau$) and exposure time ($t_{exp}$) integrated over the path from the inlet to the specific cell of interest. The equation representing this model for calculating the hemolysis index (HI) is as follows:

$$HI(\%) = \frac{dHB}{HB} \times 100 = C t_{exp}^{\alpha} \tau^{\beta} \quad (3)$$

where $dHB$ is the amount of free hemoglobin, $HB$ indicates the total hemoglobin concentration (150 g/L for adult), and $t_{exp}$ and $\tau$ are the exposure time and the scalar shear stress, respectively. Also, the values of the empirical constants $C = 1.21 \times 10^{-5}, \alpha = 0.747, and \beta = 2.004$ mentioned in the equation above have been taken from the study of Taskin et al. (Taskin et al. 2010).

It is significant that the use of animal-based blood parameters in computational models is common in the literature. Some *in-vivo* and *in-vitro* studies (Weisskopf et al. 2021; Lu et al. 2022), utilized animal models or animal-based blood to examine the performance of blood pumps.

Ding et al. ( 2015) compared hemolysis indices of human, ovine, porcine, and bovine blood under variable shear stress and exposure time conditions and found no significant difference between the blood types in the range of shear stress 50–350 Pa and exposure time of 0.05 s. In fact, the current study conditions (maximum shear stress 113.5 Pa and exposure time of slightly over 0.04 s) are consistent with these previous findings. Han et al. (2022; 2024) also agree with the use of ovine-derived parameters in such models along with using commonly used parameters (He et al. 2021), and their validity is thus established. Since the hemolysis indices are similar in all species under these conditions, the parameters selected by Taskin et al. (2010) for ovine blood are appropriate for use in our simulations.

The normalized index of hemolysis (NIH) provides an estimate of the quantitative amount of hemolysis and is expressed as:

$$NIH\ (g/100L) = 100 \times HI \times (1 - H_{ct}) \times HB \quad (4)$$

where $H_{ct}$ is the hematocrit level. Since anemia is common in patients with congestive heart failure, with a reported prevalence ranging between 15% and 56% and an average hematocrit value between 37% and 42% depending on the study (Guglin and Darbinyan 2012), we selected a value of 40% for $H_{ct}$ in this study.

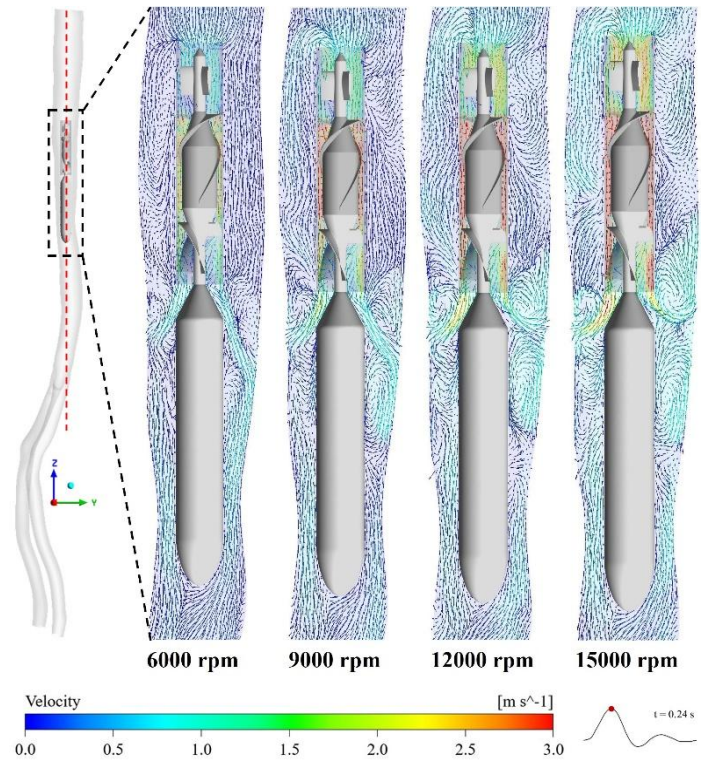

**Figure 3.** Velocity field on a mid-plane (red slice) at the peak of systolic phase inside the aorta and the pump for different rotational speeds ranging from 60000 rpm to 15000 rpm. The gray solid line is representing the cover (casing) of the pump.

## 3. Results

### *3.1. Flow field in aorta in the presence of an intra-aortic entrainment pump*

In order to examine the effect of the intra-aortic entrainment pump on the flow field in the anatomically correct aorta, velocity vectors of the flow both inside the pump and the artery were extracted from the simulations at a mid-plane at the peak of systole ($t = 0.24\ s$), as shown in **Figure 3**. When the pump is inactive, blood exhibits like laminar flow within the aorta; however, the rotational speed significantly influences the flow around the pump. Due to the high rotational speed, a jet-flow emerges at the outlet of the pump. This high-velocity flow interacts with the flow bypassing the pump and significantly impacts the region in the aorta where the pump is located. As the rotational speed of the impeller increases, the flow structures become more complex with the emergence of several vortex structures with different characteristic lengths. An interesting phenomenon observed in the aorta is the appearance of a retrograde flow, particularly evident at rotation rates higher than 9000 rpm (see the left side of the pump entrance on **Figure 3**). In addition, the interaction between the jet flow and pulsatile flow at the outlet of the axial pump starts significantly impacting the aorta wall.

The secondary flow on selected cross-sections along the aorta at the peak of the systolic phase are displayed in **Figure 4** for different pump speeds. High rotational speeds (> 6000 rpm) significantly affect the flow field in the aorta. Vortex structures of different sizes appear along the aorta. Note the small vortices appearing around the cover at plane C and the complex flow structures appearing at the exit of the pump (plane D) and further downstream (plane E).

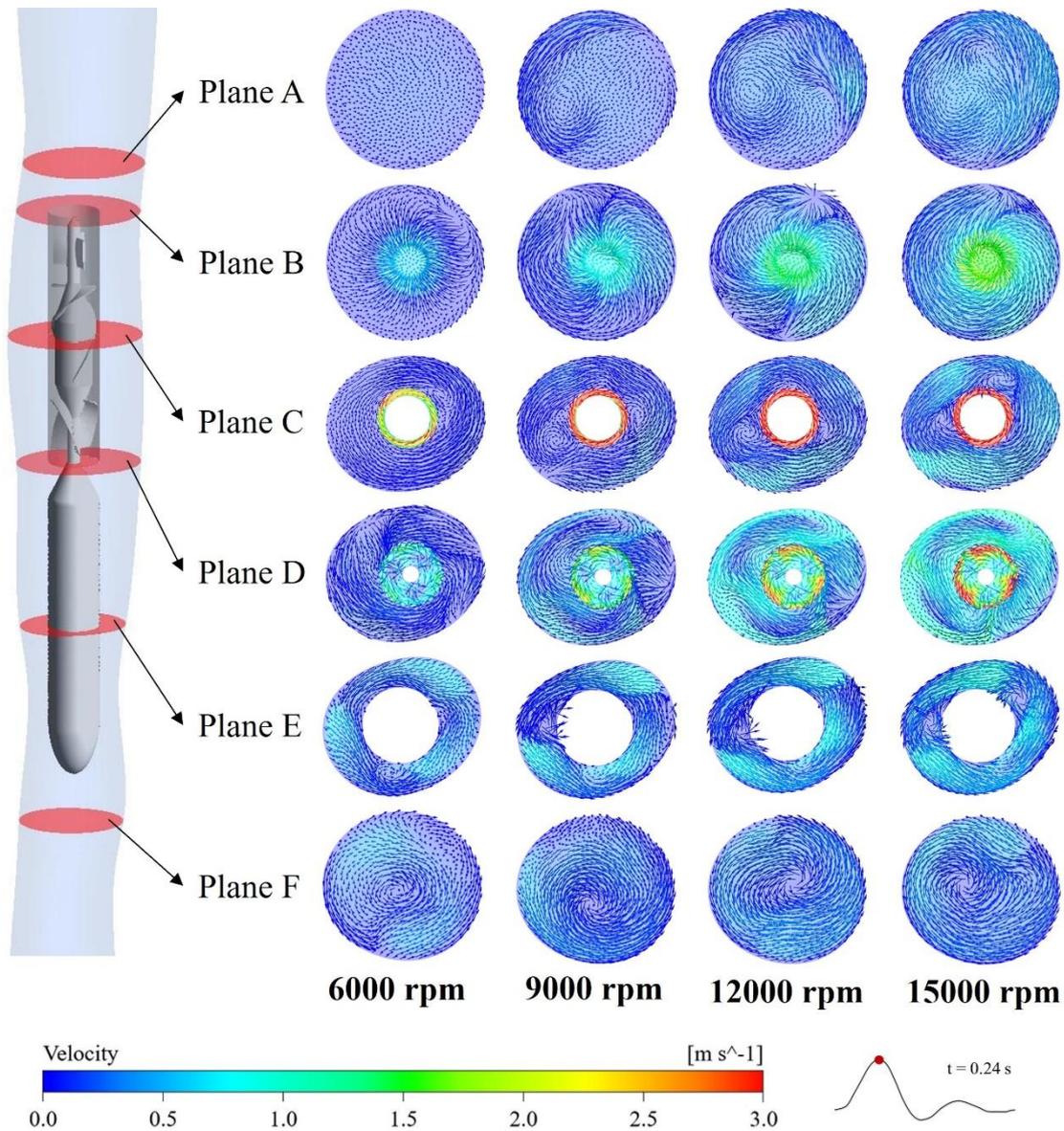

**Figure 4.** Velocity fields on different selected cross sections along the aorta.

On the cross-sectional planes through the aorta shown in **Figure 4**, in addition to the velocity fields, contour plots of the vorticity field are also shown in **Figure 5** at the peak of the systolic phase. At the inlet and outlet cross-sectional areas of the pump and inside the pump cover, the flow reaches high vorticity values. Note the evaluated values of the vorticity specifically at sections D and E corresponding to the pump exist and mid-motor section length. This is due to the pump outflow jet and its complex interaction with residual pulsatile flow and the aortic wall.

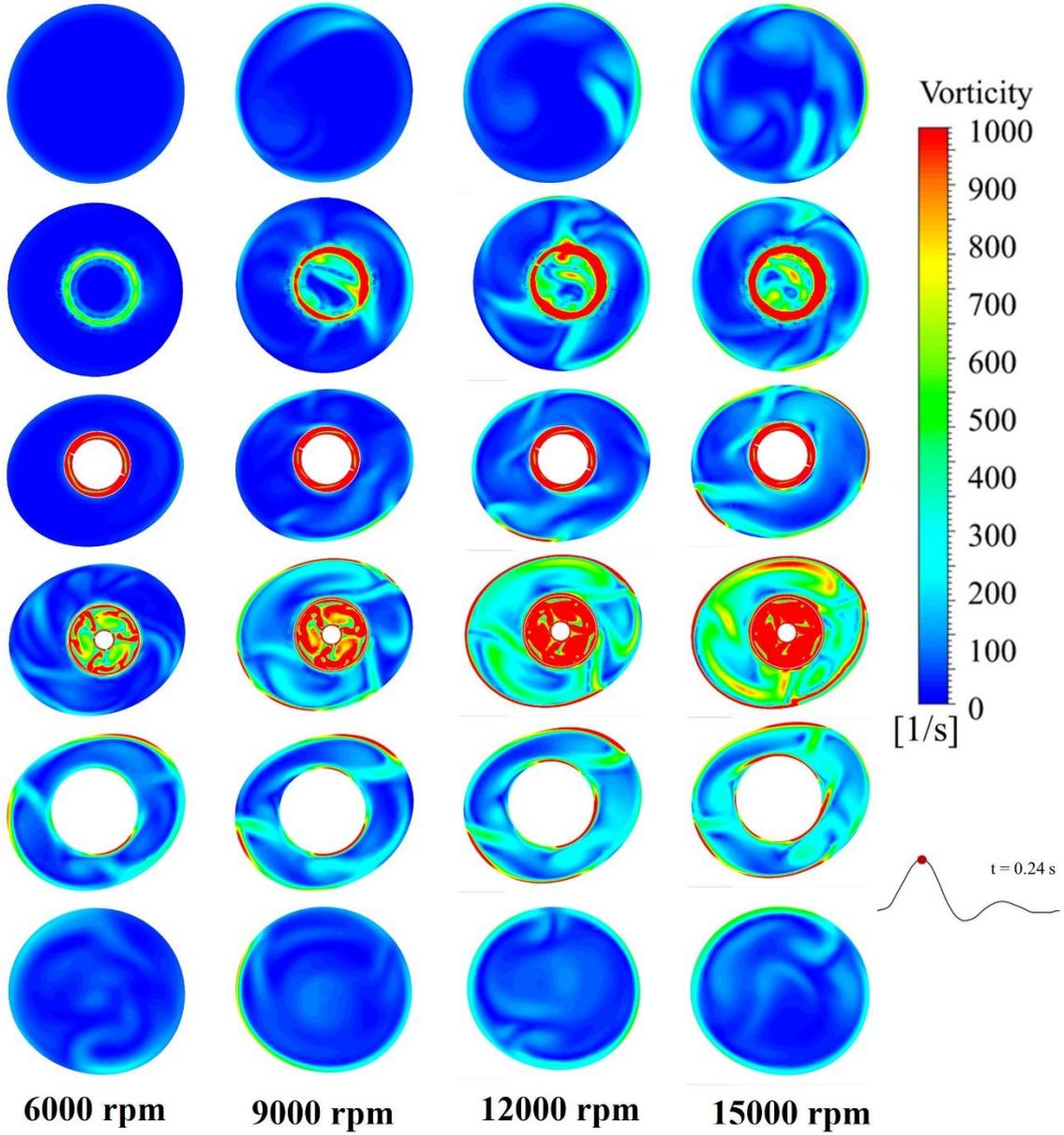

**Figure 5.** Vorticity on the selected cross sections along the aorta.

## 3.2. Wall shear stress

The distribution of wall shear stress (WSS) over the pump surface and on the aorta is a critical parameter for evaluating biocompatibility, pump efficiency, risks of atherosclerosis and vascular diseases like the development of aneurysms. **Figure 6** displays the time-averaged wall shear stress (TAWSS) over the pump surface. Time-averaged wall shear stress (TAWSS) distribution was calculated over the last cardiac cycle $T$ using Equation 5.

$$TAWSS = \frac{1}{T}\int_0^T \left(|WSS_x| + |WSS_y| + |WSS_z|\right) dt \qquad (5)$$

where, $WSS_x$, $WSS_y$ and $WSS_z$ represent the WSS values resolved in the x, y, and z directions, respectively. The baseline value, just due to the presence of the pump is very low ~ 1 Pa. Increasing the speed of the impeller lead to a significant increase in the TAWSS. The highest and most critical values are observed over the surface of the impeller compared to the straightener and diffuser, except for the blade tip of the diffuser. This part is subjected to high velocity and swirling blood flow originating from the rotating domain.

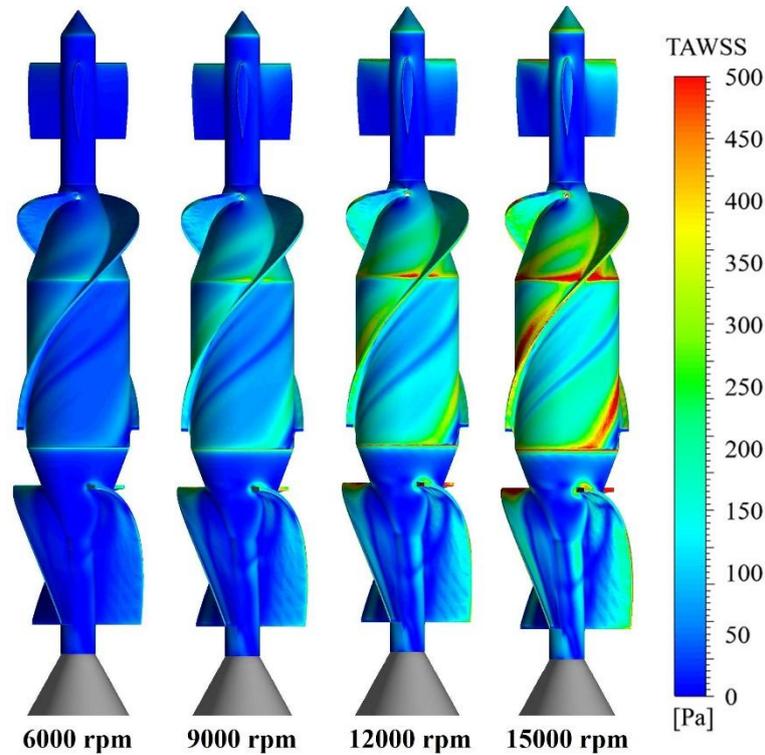

**Figure 6.** Time-averaged wall shear stress distribution over the pump surface at different rotation speeds.

**Figure 7** focusses on the aortic wall and displays the effect of the blood pump and its interaction with the aortic pulsed flow on the WSS. The baseline values for the TAWSS distribution over the aorta wall in the absence of pump rotation is ~ 1.0-1.5 Pa. When rotating the pump at different speeds, the results show that the jet-like structure emerging from the pump and interacting with the residual pulsed flow bypassing the pump led to significant localized increase in the TAWSS over the aortic wall, reaching values up to 20 Pa. Long-time exposure to such high levels of shear stress might lead to damage to the arterial wall.

*3.3. Blood damage analysis*

The prediction of the rate of blood damage inside of the pump cover has been assessed by considering HI and NIH values, as mentioned in the methods section. Figure 8 shows the variation of both HI and NIH with four different rotational speeds of pump. The value of NIH exceeds the allowable NIH line (0.1 g / 100 L) (Fu et al. 2021) as the pump rotational speed reaches 12000 rpm. More specifically, the HI increases from 0.00018% to 0.00696% for rotational speed ranging from 6000 rpm to 15000 rpm. While the NIH increases from 0.01461 to 0.57443 g/100L for the same range of rotational speed.

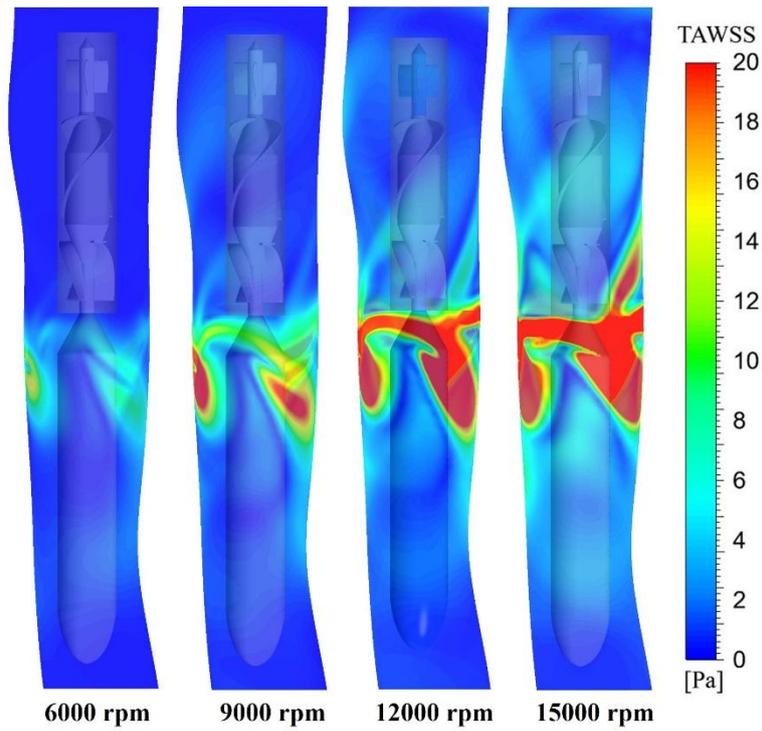

**Figure 7.** Time-averaged wall shear stress distribution over aorta wall for different rotation speeds.

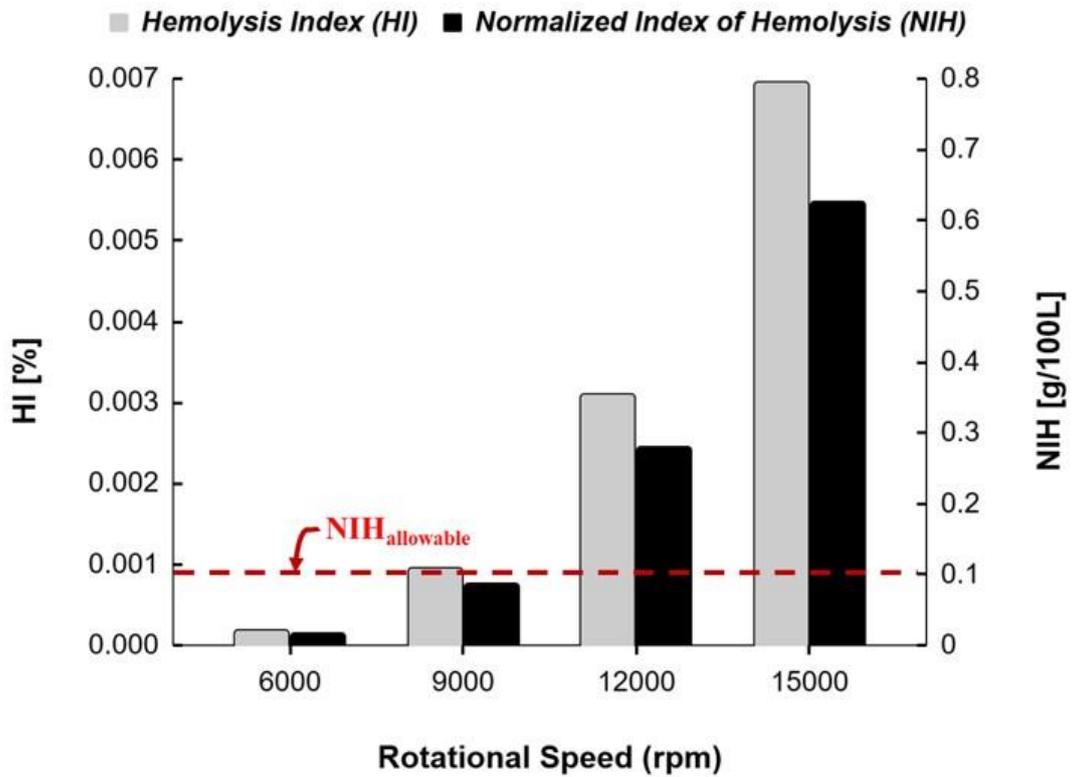

**Figure 8.** Variation of Normalized Index of Hemolysis (NIH) (Left y axis) and Hemolysis Index (HI) (right y axis) based on the rotational speeds of pump (x axis). The red dotted line indicates a value of allowable NIH, which corresponds to 0.1 g / 100 L for continuous-flow blood pumps (Fu et al. 2021).

## 4. Discussion

The main findings of this study are: 1) the intra-aortic entrainment pump leads to complex flow structures inside and outside of the pump; 2) inside the pump this leads to high levels of hemolysis even at relatively low rotational speeds; 3) outside the pump, the steady jet emerging from the pump interacts with the residual pulsatile cardiac flow and leads to elevated localized wall shear stresses on the aortic wall.

With almost six million adults in the United States experiencing heart failure, and about half of these patients having reduced ejection fraction, percutaneous mechanical circulatory support can serve as a bridge to more definitive therapies, especially among patients with advanced cardiorenal dysfunction. Early designs focused on supporting cardiac function by positioning the device as proximal as possible to the failing heart. Although successful in unloading the heart, such devices always carried the potential for elevated risks of flow-induced stroke. A new generation of devices has brought about a paradigm shift in the design of LVADs, with miniaturized, micro-axial fluid entrainment pumps mounted within a self-expanding nitinol strut anchoring system. One unique feature of such designs is their positioning in the descending aorta, meaning the device does not interact with the native aortic valve and function, minimizing the risk of thromboembolic stroke. In this configuration, a portion of the native blood flow passes through the pump and exits as a high-velocity jet, conceptually leading to an entrainment of flow that bypasses the pump. Such devices can provide partial circulatory support up to 3.5 liters per minute at nominal speeds with significant increase in renal artery blood flow and pressure (Tang et al. 2023).

In this study, the hemodynamic performance and blood damage analysis of a model of a continuous intra-aortic entrainment pump within a patient-specific aorta geometry has been examined by using computational fluid dynamics. In contrast to the only previous study dealing with the numerical evaluation of intra-aortic entrainment pumps (Li et al. 2022) in which no flow emerged from the heart, our study investigates a more realistic scenario where a residual pulsatile flow still emerges from the heart, as expected in patients with heart failure and patients who are candidates for intra-aortic entrainment pump implantation. Additionally, we focus not only on the pump and the aortic hemodynamics but also on the impact on the aortic wall. The results show that the rotational speed of the pump plays a crucial role in shaping the blood flow pattern both inside and outside the pump. The blood damage within the pump is likely related to the flow fields (Liu et al. 2020). At the maximum rotational pump speed simulated in this study, blood velocity exceeds 3 m/s, whereas the baseline value of the velocity is approximately 0.5 m/s in the case of no-rotation. The flow velocities for the blood pumps have various values, reaching higher than 4 or 5 m/s, depending on the axial pump geometry and operating speed (Throckmorton and Kishore 2009; Thamsen et al. 2015). High-speed flow resulting from rotation is frequently observed inside and at the outlet of the pump (Song et al. 2004).

The evaluation of WSS distribution over the pump and aorta surface is a significant factor in hemodynamic performance of the pump. WSS value has reached 500 Pa in some regions over the pump face at the pump speed of 12000 and 15000 rpm. The distribution of this parameter in a study about LVADs has shown similar behavior for the axial pump at a rotation speed of 9100 rpm (Romanova et al. 2022). A unique feature of this study is the investigation on the effect of the high-speed blood flow on the aorta wall has not been investigated in the literature. According to this study, WSS is significantly higher than 20 Pa on the inner surface of the aorta near the pump outlet at rotational speeds of 12000 and 15000 rpm. Long-term exposure to high level of shear stress on the aorta wall may trigger a wall response leading to the development of an aneurysm in the region surrounding the axial pump.

Elevated time-averaged wall shear stress has been shown to be responsible for arterial wall damage (Hegde et al. 2021). In the context of intra-aortic pumps, the outflow directly impinges on the aortic wall, and complex oscillatory flow structures appear around the pump. These complex structures are responsible for increased oscillatory shear stress and might contribute to the development of pseudoaneurysms. We believe this is an important issue to investigate and address in future designs. Notably, one patient (out of N=18) in the Aortix pilot study developed a pseudoaneurysm observed at the 30-day follow-up, which required surgical repair.

The analysis of blood damage has been performed using the HI and NIH as parameters. In some studies, NIH values have been obtained by conducting in-vitro experiments (Khan et al. 2018; Berk et al. 2019). Additionally, CFD predictions have been performed to assess the distortion of the red blood cells (Song et al. 2004). The findings have a wide range, mostly below the acceptable NIH value, particularly around 0.001 to 0.01 g/100L. The assessment of the hemolytic performance of the pump in this study is meaningful in comparison to the allowable limit of NIH. The pump design proves less effective for rotational speeds higher than 9000 rpm.

Finally, as stated by Weisskopf et al. (2021), one of the challenges in the design and development of actual blood pumps is, among others, to reduce the size of the device to reduce the trauma of implementation without increasing shear stresses caused by the very high rotational speeds required to deliver sufficient blood flow. In our study, we are highlighting another challenge, which is linked to considering the interaction of the device with its implementation environment and mostly specifically, in the context of our study, its impact on the arterial wall due to the high-speed jet emerging from the device.

*4.1. Limitations*

When examining the hemolysis index, the discrete phase model (DPM) can be included in the analysis to calculate the exposure time of blood cells more precisely. However, this approach can significantly increase the computational costs and time. Therefore, a simplified method has been utilized as explained in Section 2.5.

5° rotation of the pump is a small enough time step to meet the requirements for the number of wings and the wing transition frequency (Blum et al. 2022). LES might be preferred for a detailed analysis, which leads to expensive computational cost. Therefore, RANS was selected for the method in this paper.

For each cardiac period, a certain time-dependent mass flow rate is defined as the inlet boundary condition. This study focused on the hemodynamics of blood flow and the effect of jet flow on the aortic wall. In future studies, the pump characteristics should be investigated with the relationship between pump rotational speed and flow rate.

*4.2. Clinical challenges and future perspectives*

Intra-aortic pumps are focused on acute decompensated heart failure patients with complicating cardiorenal syndrome, the goal being the lowering of cardiac afterload and enhancement of renal artery blood flow (Nathan and Basir 2023). Procyrion Aortix, which was investigated in an 18-patient pilot (Cowger et al. 2023), is in trial in acute decompensated heart failure patients (NCT05677100). The major advantages of these devices are the support of the heart with less risk of stroke by positioning the pump in the descending aorta (Georges et al. 2022). The main potential issue is possible adverse effects on cerebral perfusion, i.e., 'cerebral steal.' Another study on another model (Puzzle Medical Devices) revealed a possible augmentation of cerebral flow in healthy calves, though not statistically significant. Whereas vWF activity was maintained under low pump speeds (Vincent et al. 2024). To the best of our knowledge, no study, either animal or clinical, has explicitly demonstrated stroke reduction, and more animal/clinical studies are required to establish stroke risks.

Intra-aortic pumps are designed to target patients with acute decompensated heart failure complicated by cardiorenal syndrome. Their primary objectives include reducing cardiac afterload and improving renal perfusion. For instance, devices like the Aortix can provide partial circulatory support (e.g., ~3.5 L/min) while improving renal artery blood flow and pressure by more than 35% (Nathan and Basir 2023). However, their deployment raises several clinical challenges that need careful consideration: *1) Positioning and Procedural Advantages:* The positioning of intra-aortic pumps in the descending aorta offers advantages such as reduced risk of stroke and suitability for patients with calcified or mechanical aortic valves. Unlike devices such as the Impella, which require placement across the aortic valve, intra-aortic pumps can be delivered via femoral access and are positioned downstream. However, more comparative studies are needed to assess the long-term clinical benefits of this approach; *2) Cerebral Perfusion Concerns:* A potential drawback of intra-aortic pumps is the entrainment flow mechanism, which may reduce cerebral perfusion. This phenomenon, referred to as 'cerebral steal,' has been raised as a conceptual risk and requires further clinical validation; *3) Aortic Wall and Blood Damage:* Our study highlights the potential for elevated wall shear stress and oscillatory flow structures caused by the pump's outflow impinging on the aortic wall. These stresses can damage the arterial wall, increase hemolysis risk, and potentially contribute to the development of pseudoaneurysms. Notably, one patient in the Aortix pilot study (Cowger et al. 2023) required surgical repair of a pseudoaneurysm observed at the 30-day follow-up, underscoring the clinical importance of addressing these risks in future designs. In summary, while intra-aortic pumps represent a promising innovation for managing acute decompensated heart failure, several clinical issues need further investigation, including optimizing pump design to minimize wall shear stress and blood damage, validating stroke risk reduction, and demonstrating long-term benefits in targeted patient populations. Addressing these challenges through advanced simulations and clinical trials will be essential for improving patient outcomes and device efficacy.

Our results highlight the importance of accurate simulations under real conditions, i.e. with the pump in its actual position. Most heart pump research uses straight tubes with non-pulsatile flow, whereas we simulated a patient-specific aorta with realistic pulsatile blood flow. We are developing a framework that future researchers can use to perform similar simulations and adapt to patient specifics. This strategy recognizes the risks of blood and aortic wall injury and directs the course of future designs to minimize complex interactions between pump outlet geometry and the aortic wall or pulsatile flow.

In the literature, pump rpms range from 14k to 22k for the ModulHeart (Georges et al. 2024) and 25k for the Aortix (Cowger et al. 2023). We used lower rpms in our simulations due to computational cost, but even at rpms of 15k, our results show significantly elevated risks of hemolysis and wall shear stresses on the aorta.

Intra-aortic pumps have shown promising benefits in the management of cardiorenal syndrome; however, several advancements are still needed to optimize their clinical utility and outcomes. Future research should focus on improving device design to minimize wall shear stress and reduce the potential for blood damage, which can impact patient safety and recovery. Additionally, larger-scale clinical trials are required to validate the long-term safety and efficacy of intra-aortic pumps, particularly concerning their benefits across diverse patient populations. Personalized simulations and patient-specific approaches are essential for guiding the placement and operation of these devices, allowing for potentially tailored solutions that enhance their effectiveness. Finally, as raised by (Testani et al. 2023), a deeper evaluation of the cost (required resources)-benefit of implanting a mechanical circulatory device compared to other less invasive approaches still needs to be validated.

## 5. Conclusion

Computational fluid dynamic simulations were conducted to assess the flow dynamics inside and outside a transcatheter axial pump positioned in a patient-specific abdominal aorta and subject to realistic residual pulsatile flow emanating from a failing heart. Specific attention was given to the hemolytic risks and the effect of the pump's output jet on the aorta wall. Simulations were performed for rotational speeds ranging from 6000 rpm to 15000 rpm. The results of this study show that significant blood damage may occur even for rotational speeds as low as 12000 rpm. Additionally, the interaction between the pump's outflow jet and the aortic wall leads to elevated values of wall shear stress. This needs to be carefully monitored since it might trigger the development of localized arterial aneurysms with long-term exposure. Although very promising, transcatheter axial pumps positioned in the abdominal aorta still require further improvements in their design to limit the risks of blood and aortic wall damage when used for longer periods in patients or as a bridge to transplant. Further in vivo and in vitro studies are still required to confirm the findings of the study.

**Disclosure statement**

No potential conflict of interest was reported by the authors.